\documentclass[11pt]{article}

\usepackage[a4paper,margin=1in]{geometry}

\usepackage{fontspec}
\usepackage{xeCJK}

\usepackage{indentfirst}
\IfFontExistsTF{Noto Serif CJK SC}
  {\setCJKmainfont{Noto Serif CJK SC}}
  {
    \IfFontExistsTF{Source Han Serif SC}
      {\setCJKmainfont{Source Han Serif SC}}
     
  }

\usepackage{amsmath,amssymb,amsfonts,bm,mathtools}
\usepackage{graphicx}
\usepackage{float}
\usepackage{xcolor}

\usepackage[numbers,sort&compress]{natbib}
\setcitestyle{numbers,square,comma,sort&compress}

\usepackage[colorlinks=true,linkcolor=blue,citecolor=blue,urlcolor=blue]{hyperref}
\usepackage[nameinlink,capitalise]{cleveref}

\newcommand{\dd}{\mathrm{d}}
\newcommand{\req}{\mathrm{req}}

\title{Finite-Time Thermodynamics of Battery Discharging: Power-Efficiency Trade-Off and Optimization}
\author{
Rui-Han Liu $^{1}$,
Yun-Qian Lin $^{1,*}$,
and Yu-Han Ma $^{1,2,\dagger}$\\[0.6em]
\small $^{1}$ School of Physics and Astronomy, Beijing Normal University, Beijing 100875, China\\[0.3em]
\small $^{2}$ \parbox[t]{0.86\textwidth}{\centering
Key Laboratory of Multiscale {Spin Physics (Ministry of Education),
Beijing Normal University, Beijing 100875, China}\\[0.6em]
\small $^{*}$ Correspondence: linyq@mail.bnu.edu.cn;\quad
$^{\dagger}$ Correspondence: yhma@bnu.edu.cn
}
\date{\today}
}

\begin{document}
\maketitle

\begin{abstract}
Battery discharging is governed by a fundamental trade-off between output power and energy conversion efficiency due to internal dissipation. In this paper, we demonstrate that such a trade-off universally yields a parabolic envelope $P\propto\eta(1-\eta)$. The efficiency at maximum power is exactly one half, mirroring the well-known half-Carnot limit in finite-time thermodynamics. To extend this bound into practical operational rules, we formulate a multistage constant-current discharging (MSCD) schedule subject to simultaneous real-time load demands and a global discharging deadline. Analytical resolution via the Karush--Kuhn--Tucker conditions reveals a remarkably compact optimal policy: $I_{i}^{\star}=\max(I_{i}^{-},I_{0})$. Under this rule, stages limited by external demand run exactly at their minimum required currents, while all remaining stages are elevated to a uniform baseline $I_{0}$ fixed by the deadline constraint. By tracing the dissipation--time Pareto front, we quantify how internal resistance shifts the operational boundaries and sharpens the trade-off corner. This analysis establishes a rigorous thermodynamic baseline for the scheduling layer of battery management systems, offering natural extensions to nonlinear models incorporating temperature and state-of-charge dependencies.
\end{abstract}


\section{Introduction}\label{sec:intro}

To reduce dependence on fossil fuels and lower carbon emissions, clean energy has attracted increasing attention~\cite{Liu2022}. However, the intermittent nature of renewable energy sources poses a challenge for stable energy supply~\cite{Kumar2023}. Batteries provide a practical solution to this challenge and have been widely used in electric vehicles, smartphones, and other electronic devices. Nevertheless, improper operation, such as over-charging~\cite{Kumar2023}, over-discharging~\cite{Guo2016}, and high temperatures~\cite{Maleki2006}, can degrade battery performance and compromise safety. Therefore, effective battery management is crucial for extending battery lifetime and improving operational safety and efficiency. Massive efforts have been dedicated to enhancing battery performance, encompassing novel battery designs~\cite{Grey2020,Usiskin2021,Janek2023}, advanced battery materials~\cite{liu2019,borah2020,He2023,He2024}, and optimized charging/discharging strategies~\cite{Fuller_1994,Vo2015,Lin2019}.

The core of battery operation lies in these charging and discharging cycles. In practice, Battery Management Systems (BMS)~\cite{lelie2018} schedule these cycles by tracking observable properties such as cell voltage, current, and temperature. However, existing BMS models inevitably face a practical compromise between physical precision and computational feasibility~\cite{Liu2022,Kumar2023}: detailed electrochemical models are computationally prohibitive, equivalent-circuit models require constant parameter recalibration against shifting dynamics~\cite{hu2015}, and data-driven approaches risk out-of-distribution failure. To overcome these model-dependent compromises and uncover universal performance boundaries, finite-time thermodynamics (FTT) provides a physically tractable framework~\cite{Curzon1975,Andresen1984,Bejan1996,Hoffmann1997,BejanDanMaximumWorkBattery1997,Andresen2011,tuAbstractModelsHeat2021,qiu20roadmap,zhao2025revisiting}. At its core, FTT establishes that driving energy transfer at a finite rate invariably incurs irreversible dissipation. In a battery, this manifests as a direct competition: extracting stored electrochemical energy at a larger current accelerates power delivery, but induces quadratic Joule heat across the internal resistance, thereby suppressing the discharge efficiency $\eta$~\cite{XiaMaximumWorkBattery2019}. This trade-off serves as the exact electrochemical analogue of the typical constraint relations analyzed in finite-time thermodynamic cycles~\cite{Chen1989,VandenBroeck2005,tuEfficiencyMaximumPower2008,Schmiedl2008,Esposito2010,Proesmans2016,Shiraishi2016,Pietzonka2018,maUniversalConstraintEfficiency2018,ma2021consistency,fei2022efficiency,yuanOptimizingThermodynamicCycles2022,maMinimalEnergyCost2022,zhou2024finite,zhaoEngineeringRatchetbasedParticle2024,2025MaTO,Wang2025AM}. 

Recently, two of the present authors and their collaborators successfully extended the FTT perspective to the battery charging~\cite{Lei2025}. By deriving the exact thermodynamic power-efficiency trade-off for RC circuits, they provided a theoretical foundation for empirical multistage constant-current (MSCC) protocols~\cite{Notten2005,Zhang2014MSCC,Lee2020Optimal} and determined optimal schedules through time minimization. A corresponding FTT analysis for the discharging process, however, remains absent. Unlike charging, discharging cannot be treated as a simple time-reversed step. While a charger actively tunes the input current to suppress heat, discharging is strictly load-driven~\cite{Banguero2018} and tightly constrained by external power demands alongside instantaneous voltage, current, and state-of-charge (SOC) limits~\cite{Chiasserini2001}. Delivering high power requires large currents, which inevitably exacerbate internal dissipation and spatial nonuniformity, driving the cell prematurely to its cutoff potential and shrinking its usable capacity~\cite{Masakure2023}. Furthermore, repeated high-current pulses accelerate internal resistance growth and battery degradation~\cite{matthieu2017,DehghaniSanij2019}. Since practical BMS track these degradation states through the aforementioned model approximations, the fundamental thermodynamic boundary governing the discharging process itself remains undefined. 

In this paper, we bridge this gap by applying the FTT approach to battery discharging. To unveil the underlying physical constraints, we address two core questions: i) does a battery discharging under a time-varying load obey a universal thermodynamic bound? ii) can this bound yield a closed-form optimal scheduling rule? This paper is organized as follows. In \Cref{sec:emf}, we examine three discharging models of increasing realism, encompassing constant electromotive force, finite-capacitance RC dynamics, and active constant-current control. Across all three models, we show that battery discharging obeys the same parabolic envelope $P\propto\eta(1-\eta)$, with the efficiency at maximum power equal to exactly one half. We then compare active constant-current (CC) discharge with passive resistive discharge at the same voltage window and total discharge time, showing that active control gives a lower internal heat loss. Next, \Cref{sec:mscd} introduces a multistage constant-current discharging (MSCD) schedule and solves the optimization analytically using the Karush--Kuhn--Tucker (KKT) conditions. The optimum has a simple structure: demand-limited stages operate at their minimum required currents, while the remaining stages share a uniform baseline current set by the global deadline. We then characterize the feasible operating regimes imposed by efficiency and load-power bounds, construct the dissipation--time ($Q$--$\tau$) Pareto front, and extend the analysis to the internal-resistance parameter space. Finally, \Cref{sec:outlook} concludes the results and discusses possible extensions.

\section{Power--Efficiency Parabola of Battery~Discharging}\label{sec:emf}
\unskip

\subsection{Infinite-Capacitance Regime: Constant Electromotive Force and Matched-Load~Limit}\label{sec:emf-const}

\begin{figure}[H]
  \includegraphics[width=\textwidth]{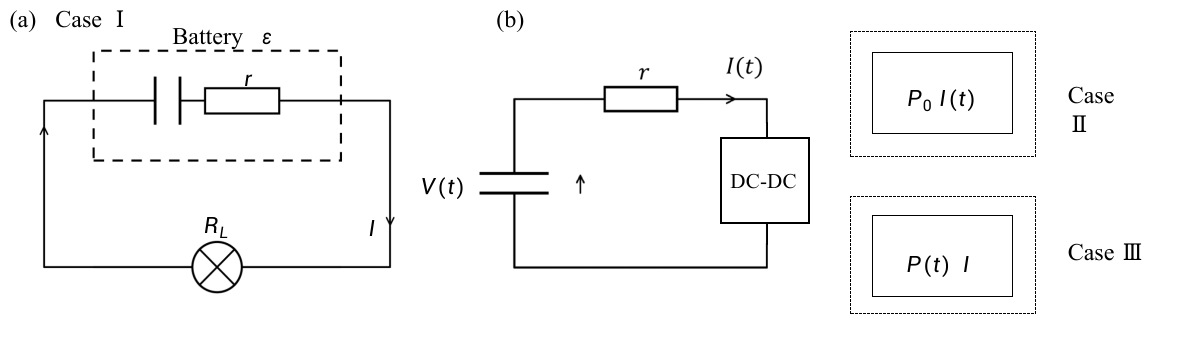}
 \caption{{Equivalent-circuit} 
schematics for the discharge models. (\textbf{a}) The constant electromotive force model, with~an ideal source in series with internal resistance $r$ driving a resistive load $R_L$. Middle: The finite-capacitance model, where a capacitor with internal resistance $r$  is coupled to a DC-DC converter that regulates the cell current $I(t)$. (\textbf{b}) The prescribed external conditions used in the analysis, namely fixed output power $P_0$ or fixed current $I$.
}
  \label{fig:circuit}
\end{figure}

{We begin with the simplest discharging model, as~illustrated in Figure~\ref{fig:circuit}a. The~battery is modeled as an effectively infinite capacitance in series with an internal resistance $r$, ensuring that its internal electromotive force $\varepsilon$ remains strictly constant throughout the entire discharging process. When supplying an external resistive load $R_L$, Kirchhoff's voltage law and energy conservation dictate that the load power is given by}

\vspace{-9pt} 

\begin{equation}
    P(I)=\varepsilon I-I^{2}r,\label{eq:power}
\end{equation}
where $\varepsilon I$ is the total chemical power of the battery and~$I^{2}r$ is the dissipative power due to circuit heating. Thus,  the~discharging efficiency is the ratio of load power to the total chemical power supplied by the battery, namely
\begin{equation}
    \eta=P/(\varepsilon I)=1-Ir/\varepsilon.\label{eq:eta}
\end{equation}
{Combining} 
Equations~\eqref{eq:power} and \eqref{eq:eta} and eliminating $I$ yields
\begin{equation}\label{eq:emf-parabola}
P(\eta)=\frac{\varepsilon^{2}}{r}\,\eta(1-\eta). 
\end{equation}
{Equation} \eqref{eq:emf-parabola} shows that the output power is a concave parabola as a function of efficiency $\eta$, vanishing at the no-load point $(\eta=1,I=0)$ and the short-circuit point $(\eta=0,I=\varepsilon/r)$. The~efficiency at maximum power  and the corresponding peak power $P_{\max}$ are achieved when $\partial P/\partial \eta=0$, yielding
\begin{equation}\label{eq:emp}
\eta_{\mathrm{EMP}}=\frac{1}{2},\qquad P_{\max}=\frac{\varepsilon^{2}}{4r}.
\end{equation} 
$P_{\max}$ is the largest instantaneous power that the battery can supply to any external load. It is attained when the load resistance is exactly the internal resistance $r$, with~the corresponding optimal current being $I^{\star}=\varepsilon/(2r)$. At~the EMP point, exactly one half of the chemical power flux is dissipated as Joule heat; this fraction is independent of the electrochemistry and is the minimum dissipation the battery incurs. 
The battery result $\eta_{\mathrm{EMP}}=1/2$ has the same structure as the linear-response result for heat engines, where the efficiency at maximum power is one half of the Carnot bound, $\eta_{\mathrm{EMP}}=\eta_C/2$~\cite{VandenBroeck2005,Esposito2010}. When operating at maximum power, half of the energy flux is inevitably dissipated as Joule heat, which can strongly affect battery performance by promoting thermal runaway and accelerating aging under high discharge currents~\cite{Bandhauer2011,Vetter2005}.

\subsection{Finite-Capacitance Regime: Two-Branch Structure and Instantaneous Power~Limit}\label{sec:branches}
In \cref{sec:emf-const}, the~current is strictly constrained by $\varepsilon$ and the load resistance $R_{L}$, namely $I=\varepsilon/(r+R_{L})$.
In practice, the~external device usually requires a prescribed  power $P_0$. To~meet this demand, the~battery is typically interfaced with the device through a power-electronic converter, such as a DC--DC converter, inverter, or~motor controller, which actively regulates the current $I(t)$ drawn from the~battery.

Therefore, we replace the constant source $\varepsilon$ by a finite capacitor $C$ with voltage $V(t)$ decreasing from $V(0)=V_{0}$ (Figure~\ref{fig:circuit}b). For~a fixed output power $P_{0}$, the~instantaneous power balance and charge conservation are
\begin{equation}\label{eq:rc-balance} V(t)I(t)=P(t)+rI^{2}(t),\qquad
I(t)=-C\frac{\dd V(t)}{\dd t}, \end{equation} 
where $r$ is the internal resistance of the battery. With~$P(t)=P_{0}$, a~real solution of the power balance in Equation~\eqref{eq:rc-balance} exists only when $V^{2}-4rP_{0}\ge 0$, setting the instantaneous power limit
\begin{equation}\label{eq:Pmax-instant} P_{0}\le
P_{\max}(t)=\frac{V^{2}(t)}{4r}, \end{equation}
taking the same form as $P_{\mathrm{max}}$ in Equation~\eqref{eq:emp}. For~$0<P_{0}<P_{\max}(t)$, the~solution for the current is given by
\begin{equation}\label{eq:Ipm} I^{\pm}(t)=\frac{V(t)\pm\sqrt{V^{2}(t)-4rP_{0}}}{2r}.
\end{equation} 
{The} instantaneous discharging efficiency $\eta=P/(VI)=1-rI/V$ follows directly from  Equation~\eqref{eq:rc-balance}. Substituting $I_{\pm}$ into this relation yields
\begin{equation}\label{eq:eta-pm}
\eta^{\mp}(t)=\frac{1\pm\sqrt{1-p}}{2},\qquad
p\equiv\frac{P_{0}}{P_{\max}(t)}=\frac{4rP_{0}}{V^{2}(t)},
\end{equation}
where $p\in[0,1]$ is the dimensionless instantaneous power. Inverting the efficiency reveals that
\begin{equation}\label{eq:dimless-p} p=4\eta(1-\eta),
\end{equation}
which is identical in form to Equation~\eqref{eq:emf-parabola}. Furthermore, the~low-current branch $I^{-}$ has $\eta^{+}>1/2$, while the high-current branch $I^{+}$ has $\eta^{-}<1/2$. Although~the two current branches can supply the load power, the low-current branch is preferred because it dissipates less and operates at a lower C-rate, both favorable for cycle life~\cite{Vetter2005,Tomaszewska2019}. At~$P_{0}=P_{\max}(t)$ the branches merge at $I=V/(2r)$. Since $P_{\max}(t)\propto V^{2}(t)$, the~deliverable peak power decreases with SOC. {It is worth emphasizing that $\eta_{\mathrm{EMP}}=1/2$ refers to the maximum-power efficiency of the idealized uncontrolled circuit shown in Figure~\ref{fig:circuit}. In~practical discharge, battery management strategies usually impose additional constraints, such as limits on the current~\cite{Plett2015} and voltage drop~\cite{PARK2023}. These constraints typically keep the battery on the high-efficiency, low-current branch, rather than operating it exactly at the maximum-power point.}

\subsection{Constant-Current Discharging and Universality of the~Parabola}\label{sec:const-I}

Constant-current (CC) discharging is a basic strategy in battery use and BMS capacity testing~\cite{Kumar2023,PARK2023}, which can be implemented by converter control. Here, we consider a finite battery with a constant-current demand undergoing a full discharging process from  $V_{0}$ to the cut potential $V_{f}$~\cite{besenhard2008handbook}. With~charge conservation, the~total discharge time $t_{f}=C(V_{0}-V_{f})/I$ and the instantaneous potential $V(t)=V_{0}-(I/C)t$ are~maintained.

Moreover, the~released free energy during discharging follows $\Delta E= {C(V_{0}^{2}-V_{f}^{2})}/{2}$, while the cumulative Joule heat is $Q=rIC(V_{0}-V_{f})$.  According to the first law of thermodynamics, $W_{\mathrm{out}}=\Delta E-Q$ is obtained as
\begin{equation}
    W_\mathrm{out}=C(V_0 - V_f)\left( \frac{V_0 + V_f}{2} - rI \right).
\end{equation}
{To} quantify the thermodynamic performance of the discharging process, we introduce the discharging efficiency $\overline{\eta}\equiv{W_{\mathrm{out}}}/{\Delta E}$ and discharging power $\overline P=W_{\mathrm{out}}/t_f$, which, respectively, read
\begin{equation}\label{eq:bareta-const-I}
\overline{\eta}\equiv1-\frac{rI}{\overline  V},\qquad \overline P(I)=C\frac{V_0 - V_f}{t_f} \left( \frac{V_0 + V_f}{2} - rI \right),
\end{equation}
where $V\equiv\frac{V_{0}+V_{f}}{2}$. Substituting  $I=(\overline V/r)(1-\overline\eta)$ into $\overline P(I)$ gives
\begin{equation}\label{eq:Pbar-const-I}
\overline P(\overline\eta)=\frac{\overline V^{2}}{r}\,\overline\eta(1-\overline\eta).
\end{equation}
{Thus,} the CC process has the same parabola as Equation~\eqref{eq:emf-parabola}, with~$\varepsilon$ replaced by the mean open-circuit voltage $\overline V=(V_0+V_f)/2$. 

\begin{figure}[H]
  \includegraphics[width=\textwidth]{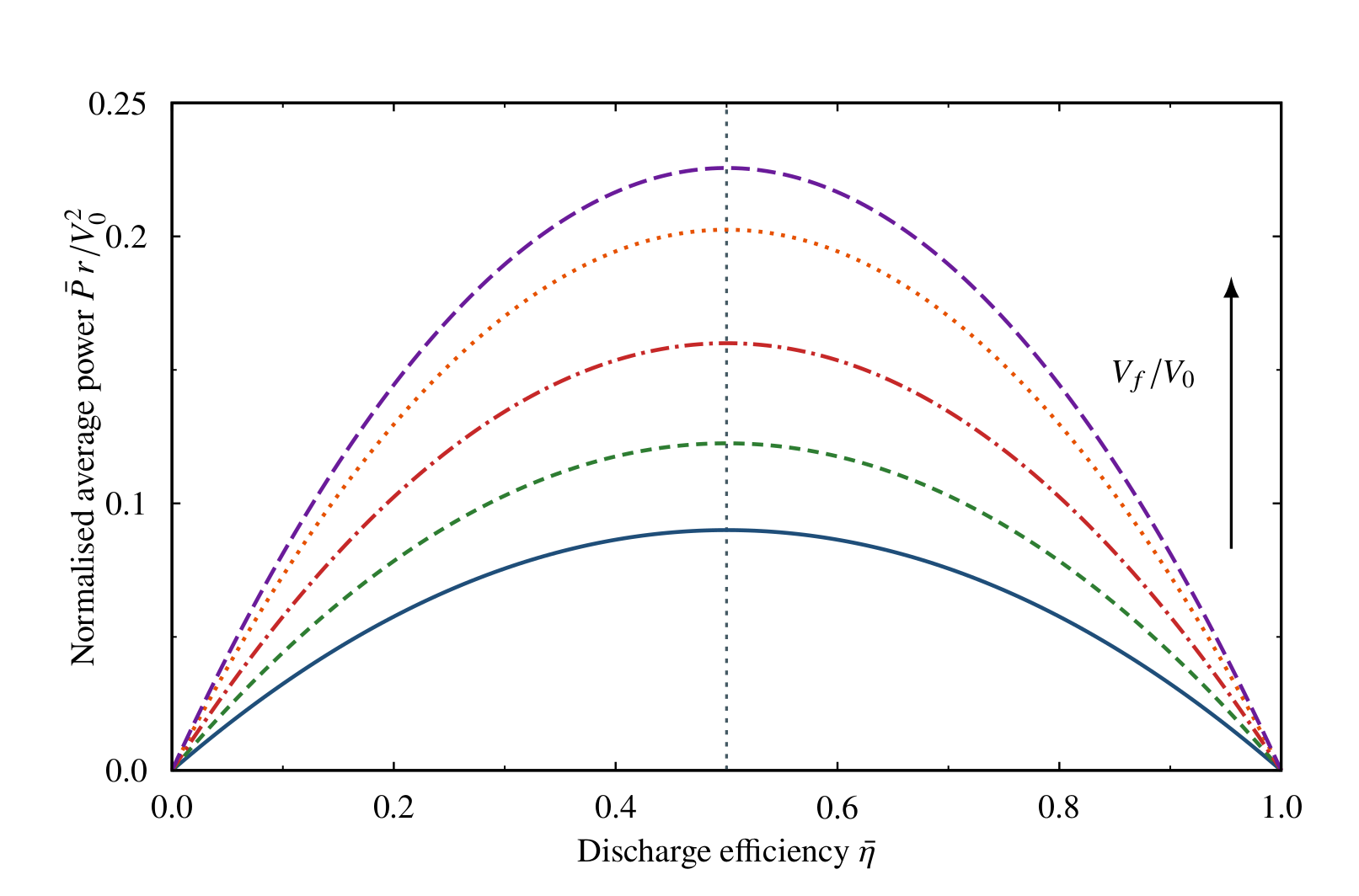}
  \caption{Normalized average power $\overline P\,r/V_{0}^{2}$ versus process-averaged efficiency $\overline\eta$ for constant-current discharging at five cut-off ratios from deep to shallow discharge. From~bottom to top, the~curves correspond to $V_f/V_0=0.20,0.40,0.60,0.80,$ and $0.90$. The~dotted vertical line marks the maximum-power efficiency $\overline\eta=1/2$.}
  \label{fig:cutoff}
\end{figure}

Figure \ref{fig:cutoff} plots Equation~\eqref{eq:Pbar-const-I} with a series of cut-off ratios $V_f/V_0$. It is obvious that the curves exhibit the same parabolic dependence on $\overline\eta$: the curves vanish at $\overline\eta=0$ and $1$, and~peak at $\overline\eta=1/2$.  As~$V_f/V_0$ increases, the~curve is lifted due to the larger vertical scale $\overline V^{2}/r$. Therefore, a~shallower discharge permits a higher average power at a fixed efficiency, which is consistent with the performance improvement observed in pulsed-discharge protocols~\cite{Lv2020}. In~the limit $V_f/V_0\to1$, the~peak approaches the constant-EMF value with $\varepsilon=V_0$.

The emergence of the identical parabolic envelope $P \propto \eta(1-\eta)$ across all three models discussed in the current section reveals a robust thermodynamic universality. It establishes that the efficiency at maximum power is exactly $\eta=1/2$. Because~the theoretical reversible limit for battery discharging is unity ($\eta_{\mathrm{rev}}=1$), this optimum represents exactly half of the reversible bound, mapping rigorously to the well-known $\eta_{C}/2$ efficiency at maximum power (the half-Carnot limit) in finite-time heat engines~\cite{VandenBroeck2005,Esposito2010,tuEfficiencyMaximumPower2008,Proesmans2016,zhao2025revisiting,qiu20roadmap}. This yields a fundamental takeaway: while internal chemistry and external control of the battery dictate the absolute power scale, the~normalized parabolic envelope remains an invariant signature of Ohmic~dissipation.

{We close this section with a remark on the scope and limitations of the foundational model. The~equivalent-circuit model adopted here serves as a minimal analytical baseline to derive the thermodynamic bounds and formulate the MSCD optimization. By~using a finite capacitance to represent the SOC-dependent open-circuit voltage, the~model captures the inherent trade-off between average power and extractable energy~\cite{Tsirlin2024}. While real batteries involve complex polarization, diffusion, and~entropy--heat dynamics that necessitate higher-order or electrochemical models~\cite{Hu2012,Plett2015,Doyle1993,Newman1975,Sulzer2021,Yang2025Energy139041}, we deliberately isolate the primary Ohmic-loss mechanism responsible for the parabolic envelope, leaving these advanced integrations for future work. Furthermore, this baseline temporarily omits series inductance. Because~the MSCD protocol operates via piecewise constant-current stages ($dI/dt \approx 0$), inductive effects strictly vanish during bulk discharging and only perturb the brief switching transients without altering the stage-wise Joule dissipation. A~rigorous discussion of these inductive effects is deferred to Section~\ref{sec:inductance-summary}.}

\subsection{Performance of CC~Discharging}\label{sec:active-knee}

In the passive circuit, no converter is present to regulate the current or output power. The~battery simply discharges through a fixed external load $R_L$, leading to the uncontrolled RC relaxation
\begin{equation}
V(t)=V_0 e^{-t/[(r+R_L)C]},
\end{equation}
where the capacitor is finite. By~contrast, in~CC discharge, a~power-electronic controller is introduced to enforce a prescribed current profile. Here, we compare the performance of CC discharging with passive discharging under a fixed total discharge time $t_f$ over the same voltage window $[V_f,V_0]$.

\begin{figure}[H]
\hspace{-9pt}
  \includegraphics[width=9.2cm]{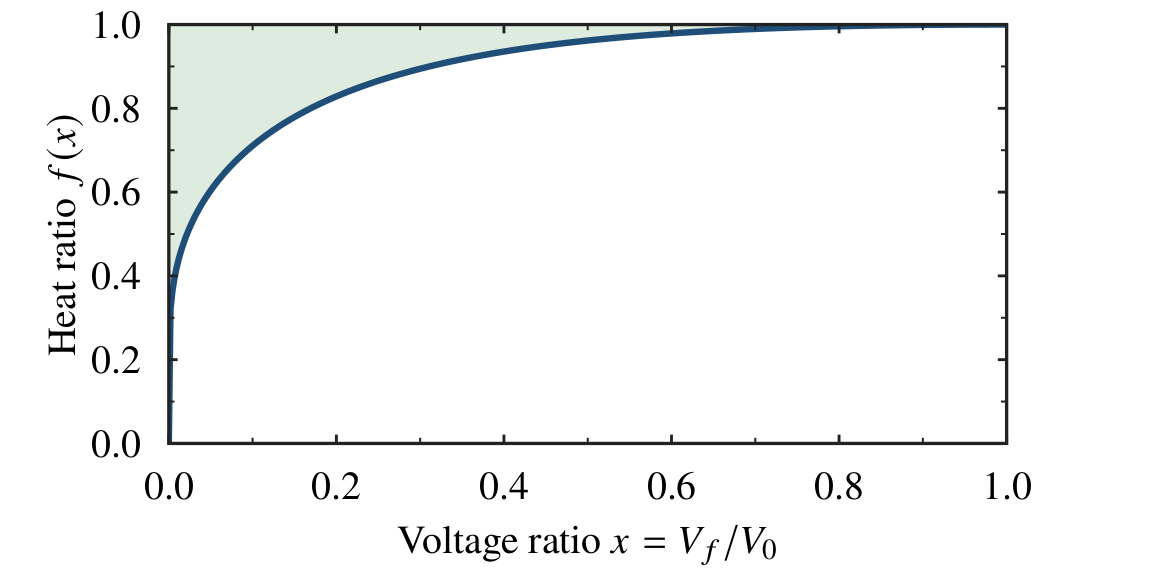}
  \caption{Heat ratio $f(x)=Q_{\mathrm{diss,active}}/Q_{\mathrm{diss,passive}}$ is the function of the voltage ratio $x=V_{f}/V_{0}$. The~green region is the saving fraction of dissipation $1-f$. }
  \label{fig:active-passive}
\end{figure}

Fixing $V_f$ yields the total resistance in the circuit $r+R_{L}=t_{f}/[C\ln(V_{0}/V_{f})]$, with~which the dissipation induced by the internal resistance $r$ is given by
\begin{equation}
\label{eq:Q-passive}
Q_{\mathrm{diss,passive}} =\tfrac{1}{2}C(V_{0}^{2}-V_{f}^{2})\,
\frac{rC\ln(V_{0}/V_{f})}{t_{f}}. 
\end{equation}
{Under} the constraint of charge $C(V_{0}-V_{f})$, the~current in CC discharge is, thus, $I=C(V_{0}-V_{f})/t_{f}$, revealing that the internal dissipation
\begin{equation}\label{eq:Q-active}
Q_{\mathrm{diss,active}}=\frac{C^{2}r(V_{0}-V_{f})^{2}}{t_{f}}. 
\end{equation} 
{We} quantify the performance of the active control by
\begin{equation}\label{eq:fx}
f(x)\equiv\frac{Q_{\mathrm{diss,active}}}{Q_{\mathrm{diss,passive}}}
=\frac{2(1-x)}{(x+1)\ln (1/x)}, 
\end{equation}
where $x=V_f/V_0$ is the voltage ratio and $(1-x)$ means the depth of discharge in the battery. Figure~\ref{fig:active-passive} plots $f(x)$, which remains below unity over the whole range of $x$. In~other words, active CC control always dissipates less heat than passive discharge under the same voltage window and discharge~time.

\unskip

\section{Multistage Constant-Current Discharging (MSCD)}\label{sec:mscd}

\Cref{sec:branches} shows that, for~a prescribed power demand $P$, operation on the high-efficiency branch imposes a minimum admissible current $I^-$. Increasing the current can accelerate the discharging, but~only at the expense of enhanced Ohmic dissipation and reduced efficiency. Therefore, fast discharge naturally becomes a constrained optimization problem: the current should be increased to shorten the discharge time, while remaining within the admissible range set by the power demand and efficiency~constraint.

Motivated by the multistage constant-current (MSCC) protocol widely used in charging~\cite{Notten2005,Zhang2014MSCC,Lee2020Optimal}, we introduce a multistage constant-current discharging (MSCD) protocol. In~this protocol, the~discharge process is divided into stages, and~a constant current $I_i$ is assigned to each stage while satisfying the stage-wise requirement $I_i\ge I_i^{-}$. The~resulting problem is not a simple time reversal of charging: the load-power demand imposes lower bounds on the stage currents, while all stages are coupled through the global discharge-time constraint. The~optimal MSCD schedule is, therefore, obtained by solving the Karush--Kuhn--Tucker (KKT) conditions~\cite{Boyd2004,Bertsekas1999}, as~detailed in Section~\ref{sec:mscd-opt}.

\subsection{Stage-Wise Description and the Unconstrained Minimum-Dissipation~Bound}\label{sec:mscd-math}

We partition the voltage window $[V_{f},V_{0}]$ into $N$ stages by the levels $V_{0}>V_{1}>\dots>V_{N}=V_{f}$. The~initial voltage and the discharging current in the $i$th stage  are $V_{i-1}$ and $I_i$. The~discharging time $\Delta t_{i}$, the~released charge $\Delta q_{i}$, and~the stage Joule heat $Q_{i}$ are, respectively, given by:
\begin{align}
\Delta t_{i}&=\frac{C(V_{i-1}-V_{i})}{I_{i}},\\ 
\Delta q_{i}&=C(V_{i-1}-V_{i}),\label{eq:dqi}\\ Q_{i}&=I_{i}^{2}\,r\,\Delta t_{i}=rC(V_{i-1}-V_{i})I_{i}.
\end{align}
{Therefore}, the~total discharge time $\tau$ and the total dissipated Joule heat $Q$ are
\begin{align}
\tau&=\sum_{i=1}^{N}\frac{C(V_{i-1}-V_{i})}{I_{i}},\label{eq:tau}\\
Q&=\sum_{i=1}^{N}rC(V_{i-1}-V_{i})I_{i}.
\end{align}
{Supposing} only $V_{0}$, $V_{f}$ and the total time $\tau$ are prescribed and $r$ , $C$ are constants, the~Cauchy--Schwarz inequality gives $Q\tau
 \ge\bigl[rC^{2}\sum_{i}(V_{i-1}-V_{i})\bigr]^{2}=rC^{2}(V_{0}-V_{f})^{2}$.
Equality holds if all $I_{i}$ are equal, namely  a single-current discharge. This is the discrete-stage counterpart of the Salamon--Berry thermodynamic-length result~\cite{Salamon1983,Andresen2011} and sets the baseline against which the constrained optimum of Section~\ref{sec:mscd-opt} is~measured.

\subsection{Optimal Protocol  Under Power and Deadline~Constraints}\label{sec:mscd-opt}

Although the protocol is now multistage, each stage individually is still governed by the discussion in Section~\ref{sec:branches}. Evaluating the high-efficiency branch of Equation~\eqref{eq:Ipm} at $V=V_{i}$ and $P_{0}=P_{i}^{\req}$ gives
\begin{equation}\label{eq:Ireq}
I_{i}^{-}=\frac{V_{i}-\sqrt{V_{i}^{2}-4rP_{i}^{\req}}}{2r},\qquad
P_{i}^{\req}\le\frac{V_{i}^{2}}{4r}.
\end{equation}
{The} MSCD optimization problem is then
\begin{equation}\label{eq:opt-problem}
\begin{aligned}
\min_{\{I_{i}\}}\;&\sum_{i=1}^{N}rC(V_{i-1}-V_{i})I_{i},\\
\text{s.t.}\;&\sum_{i=1}^{N}\frac{C(V_{i-1}-V_{i})}{I_{i}}\le\tau_{\max},
\quad I_{i}\ge I_{i}^{-}.
\end{aligned}
\end{equation}

MSCD differs from MSCC through its stage-wise lower bound. In~the MSCC treatment of Lei~et~al.~\cite{Lei2025}, the~free stage currents are obtained by minimizing the total charging time, yielding $I_i^2=I_{i-1}I_{i+1}$ and, hence, a decreasing current sequence with equal stage time. However, in~MSCD, each stage must satisfy the lower bound $I_i\ge I_i^{-}$. Because~it is not known in advance which lower bounds will bind, the~optimum must be determined from the KKT complementary-slackness conditions, leading to the two groups in  Equation~\eqref{eq:Istar}.

To incorporate the two constraints, we introduce a Lagrange multiplier $\lambda\ge 0$ for the global discharge deadline and multipliers $\mu_{i}\ge 0$ for each lower bound $I_{i}\ge I_{i}^{-}$. The~Lagrangian is
\begin{equation}\label{eq:lagrangian}
\begin{aligned}
  \mathcal{L}
  =\,&\sum_{i}r\Delta q_{i}I_{i}
   +\lambda\!\left(\sum_{i}\frac{\Delta q_{i}}{I_{i}}-\tau_{\max}\right)
   -\sum_{i}\mu_{i}(I_{i}-I_{i}^{-}),
\end{aligned}
\end{equation}
where the stage charge $\Delta q_{i}=C(V_{i-1}-V_{i})$ from  Equation~\eqref{eq:dqi}, and the~total heat and time simplify to $Q=\sum_{i}r\Delta q_{i}I_{i}$ and $\tau=\sum_{i}\Delta q_{i}/I_{i}$.

The KKT stationarity condition applies to each stage current:$\partial\mathcal{L}/\partial I_i=0$, revealing
\begin{equation}\label{eq:foc}
I_i^2=\frac{\lambda/r}{1-\mu_i/(r\Delta q_i)} .
\end{equation}
{The} complementary-slackness condition, $\mu_i(I_i-I_i^-)=0$, encodes the active-bound structure of the lower-current constraint. For~$\mu_i=0$, the~stage shares the common current $I_0\equiv\sqrt{\lambda/r}$. When $\mu_i>0$, the~lower bound is active, enforcing $I_i=I_i^-$, namely, the stage current is fixed at its minimum admissible value.
Combining the two cases gives
\begin{equation}
\label{eq:Istar}
I_i^\star=\max\!\bigl(I_i^-,\,I_0\bigr),
\end{equation}
where $I_0\equiv\sqrt{\lambda/r}$ is shared by all unpinned stages, while pinned stages remain at $I_i^-$. With~the global discharge deadline, $I_0$ is governed by
\begin{equation}
\label{eq:I0-eq}
\sum_{i=1}^{N}\frac{C(V_{i-1}-V_i)}{\max(I_i^-,\,I_0)}
=\tau_{\max}.
\end{equation}
{Since} the left-hand side decreases monotonically with $I_0$, $I_0$ can be efficiently obtained by bisection.
Substituting $I_i^\star$ into the heat loss, we obtain
\begin{equation}
\label{eq:QMSCD}
Q_{\mathrm{MSCD}}(\tau_{\max})
=\sum_{i=1}^{N}rC(V_{i-1}-V_i)\,I_i^\star(\tau_{\max}).
\end{equation}

\begin{figure}[H]
\hspace{-3pt}
  \includegraphics[width=\textwidth]{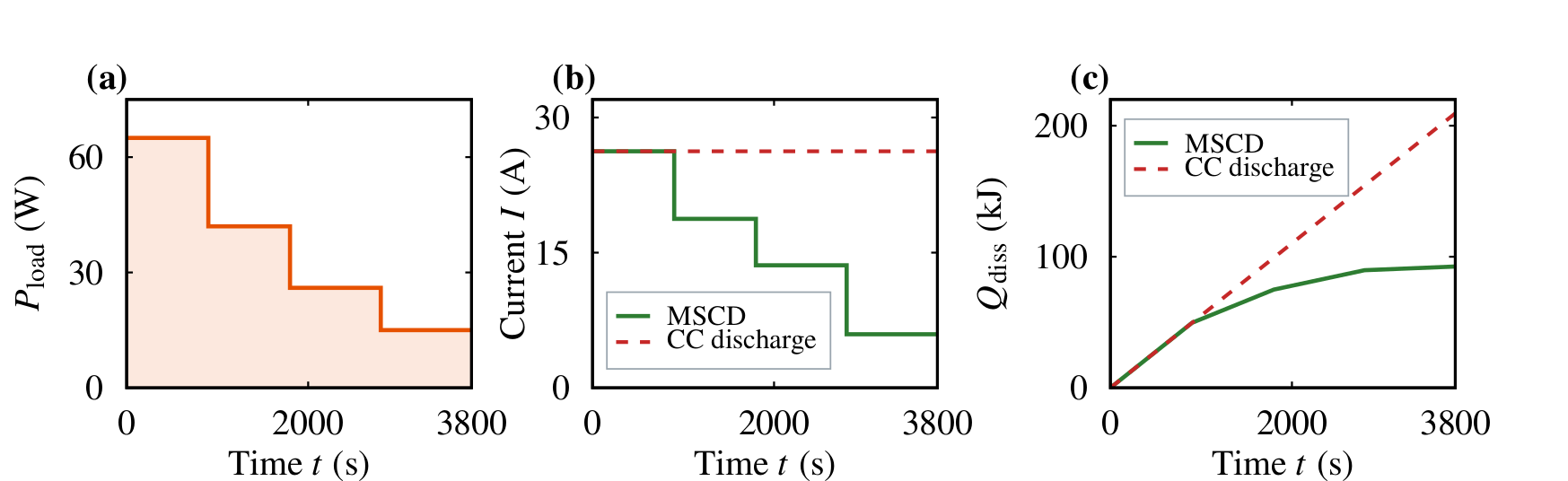}
  \caption{{Comparison of} 
 two discharging protocols under a stepwise dynamic load. (\textbf{a})~The prescribed stepwise load-power demand $P_{\mathrm{load}}(t)$ (orange shaded region). (\textbf{b})~The discharging current applied in MSCD (green solid line) and CC discharge (red dashed line). (\textbf{c})~Accumulated Joule heat $Q_{\mathrm{diss}}(t)$ during~discharging.}
  \label{fig:mscd-demo}
\end{figure}

\Cref{fig:mscd-demo} compares the optimized MSCD protocol with conventional CC discharge. The~load demand is prescribed in a stepwise manner and remains constant within each stage, as~shown in Figure~\ref{fig:mscd-demo}a. To~meet this demand, the~MSCD protocol assigns a stage-dependent constant current, whereas the CC protocol applies the peak required current throughout the entire discharge process, as~shown in Figure~\ref{fig:mscd-demo}b. The~corresponding accumulated heat losses are plotted in Figure~\ref{fig:mscd-demo}c. Compared with CC discharge, for which $Q=rIC(V_{0}-V_{f})$, the~MSCD protocol gives the stage-wise result in \cref{eq:QMSCD} and reduces the total dissipation by lowering the current with lower load demands. This result shows that adapting the current to the stage-wise load demand suppresses unnecessary Joule heating while still satisfying the required power~output.

\vspace{-12pt}

As a practical five-stage example, we use the positive discharge-power steps of the DOE/INL PHEV charge-depleting cycle-life test
profile~\cite{Belt2010PHEVTestManual}. The~system-level powers are
$P_{\mathrm{sys},i}
=\{50,45,28.125,22.5,11.25\}\,\mathrm{kW}$. Following the Battery Size Factor (BSF) scaling rule in the same manual, we take $\mathrm{BSF}=1400$ and obtain the stage load on a single battery demanding
\begin{equation}
P_i^{\req}
=\frac{P_{\mathrm{sys},i}}{\mathrm{BSF}}
=\{35.71,32.14,20.09,16.07,8.04\}\,\mathrm{W}.
\end{equation}
{We} use the representative voltage grid $(V_0,V_1,\ldots,V_5)={4.2,4.0,3.75,3.5,3.25,3.0}~\mathrm{V}$, together with an internal resistance $r=0.08~\Omega$~\cite{Hu2012,Plett2015}. The~effective capacitance is set to $C=1.103\times10^{4}~\mathrm{F}$, corresponding to an effective capacity of $3.68\mathrm{Ah}$ over the voltage window $4.2-3.0~\mathrm{V}$. For~these parameters, Equation~\eqref{eq:Ireq} gives the minimum admissible currents $I_i^-={11.64,11.29,6.80,5.76,2.90}~\mathrm{A}$. If~each stage operates at its lower bound, the~resulting load-following schedule takes $\tau_{\mathrm{req}}\simeq37.8~\mathrm{min}$. To~illustrate the deadline-constrained optimization, we impose a tighter total discharge time, $\tau_{\max}=21.94~\mathrm{min}$, corresponding to the Pareto-knee point determined later in \Cref{sec:3d}. Solving Equation~\eqref{eq:I0-eq} yields $I_0\simeq9.37\,\mathrm{A}$. Therefore, the~first two stages remain pinned to $I_i^-$, while the last three are raised to $I_0$, giving $I_i^\star=\{11.64,11.29,9.37,9.37,9.37\}\,\mathrm{A}$. Substituting these currents into Equation~\eqref{eq:QMSCD} gives $Q_{\mathrm{MSCD}}\simeq10.75\,\mathrm{kJ}$ and $\eta_{\mathrm{MSCD}}\simeq77.4\%$. Compared with the single-peak-current benchmark, which holds $\max_i I_i^-$ throughout all stages, MSCD reduces Joule heat from $12.32\,\mathrm{kJ}$ to $10.75\,\mathrm{kJ}$, corresponding to a $12.8\%$ reduction, and~raises the efficiency from $74.1\%$ to $77.4\%$.

\subsection{Feasibility Under Efficiency and Load~Bounds}\label{sec:Nfeas}
The same KKT analysis also applies when the constraint is placed on instantaneous efficiency. For~a finite-capacitance battery, when no external power demand is imposed, a~natural alternative optimization problem is to minimize dissipation over a prescribed discharge time while enforcing a lower bound $\eta_0$ on the instantaneous efficiency. In~an MSCD stage, the~instantaneous efficiency is given by
\begin{equation}
\label{eq:eta-stage}
\eta_i(V)=1-\frac{rI_i}{V}.
\end{equation}
{Since} $V$ decreases within the stage, the~minimum efficiency occurs at the stage end $V_i$. The~constraint $\eta_i(V_i)\ge\eta_0$, therefore, yields the stage-wise current upper bound
\begin{equation}
\label{eq:Ieta}
I_i\le I_i^{\eta}
=\frac{(1-\eta_0)V_i}{r}.
\end{equation}
{Repeating} the KKT analysis gives the analogous form
\begin{equation}
\label{eq:Istar-eta}
I_i^\star=\min(I_c,I_i^{\eta}),
\end{equation}
where the common current $I_c$ is governed by
\begin{equation}
\label{eq:Ic-eq}
\sum_{i=1}^{N}
\frac{C(V_{i-1}-V_i)}{\min(I_c,I_i^{\eta})}
=\tau_{\max}.
\end{equation}
{Thus}, the~efficiency constraint sets the upper bound on the stage-wise current, whereas the load-power constraint in Equation~\eqref{eq:Istar} sets the lower~bound.

For a fixed efficiency lower limit $\eta_0$, the~current upper bounds $I_i^{\eta}$ are fixed, and~changing the deadline only changes the common current $I_c$ in Equation~\eqref{eq:Ic-eq}. The~shortest feasible discharge occurs when all stages are driven at their upper bounds,
\begin{equation}
\label{eq:tau-min-eta}
\tau_{\min}^{\eta}
=
\sum_{i=1}^{N}
\frac{C(V_{i-1}-V_i)}{I_i^{\eta}} .
\end{equation}
{If} $\tau_{\max}<\tau_{\min}^{\eta}$, the~discharge is infeasible. As~$\tau_{\max}$ increases, $I_c$ decreases: just above $\tau_{\min}^{\eta}$, low-voltage stages are held at their upper bounds, while for a loose deadline, all upper bounds become inactive and the solution reduces to ordinary constant-current~discharge.

\begin{figure}[H]
  \includegraphics[width=\textwidth]{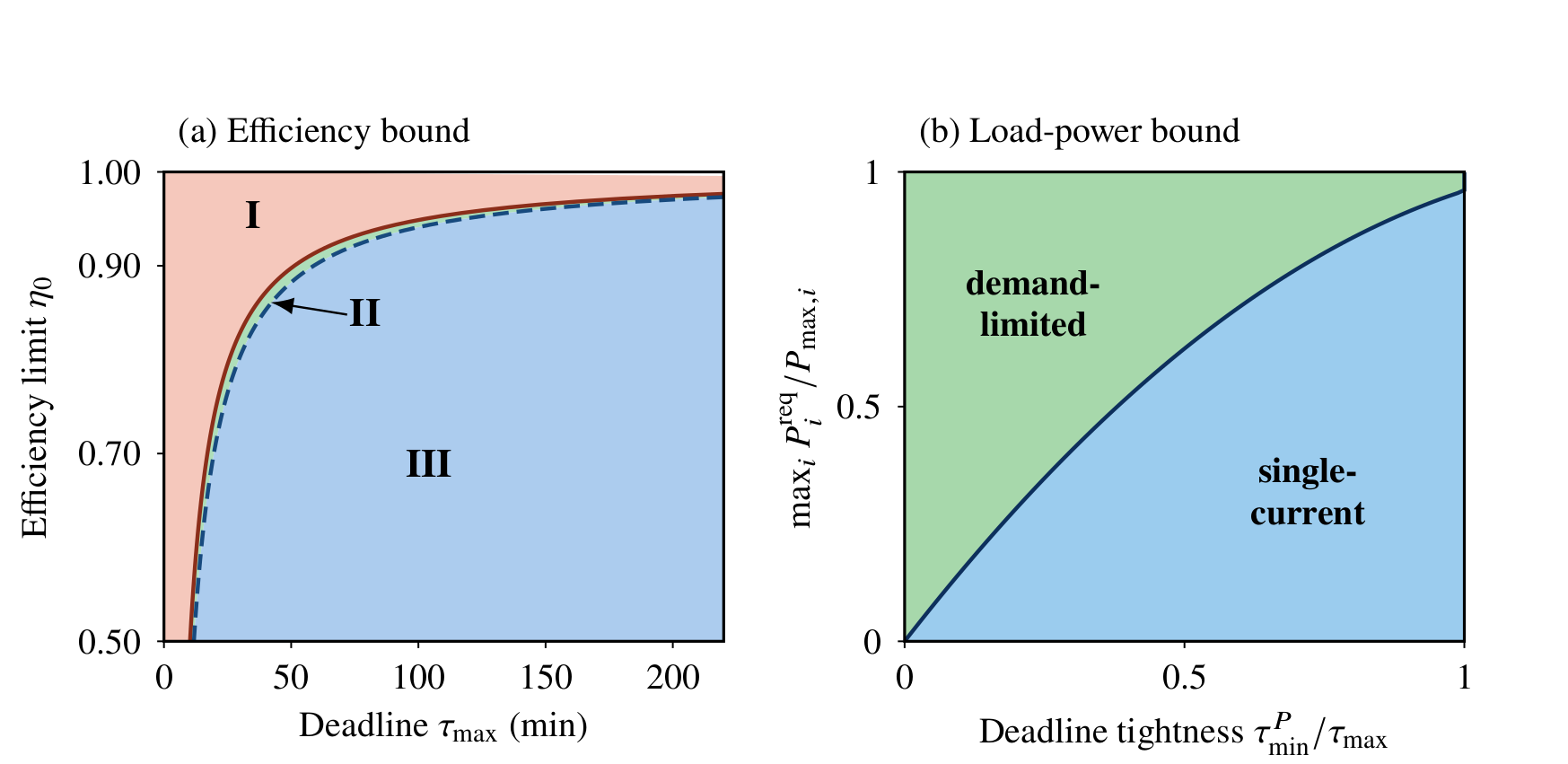}
  \caption{{Feasibility maps} 
 for the two stage-wise bounds in MSCD.
(\textbf{a}) A minimum instantaneous efficiency $\eta_0$ sets current upper bounds $I_i^\eta$. The~solid and dashed curves denote $\tau_{\max}=\tau_{\min}^{\eta}$ and $I_c=\min_i I_i^\eta$, respectively. They divide the plane into the infeasible region (I), bounded-MSCD region (II), and~CC region (III).
(\textbf{b}) In the load-power-constrained case, prescribed stage powers set current lower bounds $I_i^-$. The~axes are $\max_i P_i^{\req}/P_{\max,i}$ and
$\tau_{\min}^{P}/\tau_{\max}$, and~the curve $I_0=\max_i I_i^-$ separates the
demand-limited MSCD region from the single-current region.}
  \label{fig:feasibility-bounds}
\end{figure}

This behavior is summarized in Figure \ref{fig:feasibility-bounds}{a}, 
where the efficiency lower limit $\eta_0\in[0.5,1)$ is plotted against the deadline $\tau_{\max}$ for the high-efficiency range considered here. The~solid curve gives the minimum feasible time, $\tau_{\max}=\tau_{\min}^{\eta}$, for~a given $\eta_0$ so that region I is infeasible. The~dashed curve is the upper-bound inactivity bound, $I_c=\min_i I_i^\eta$, at~which the constrained MSCD solution reduces to CC discharge. Region II is the efficiency-bounded MSCD region between the two curves, where at least one low-voltage stage is held at its current upper bound. Region III is the CC region, where the common current is below all stage current upper~bounds.

The load-constrained case has the opposite structure, because~the prescribed stage powers set current lower bounds rather than upper bounds. For~fixed load demands $P_i^{\req}$, the~lower bounds $I_i^-$ are fixed. A~shorter deadline requires a larger common current $I_0$, whereas a longer deadline allows a smaller $I_0$. The~shortest feasible time is obtained when every stage is driven at the maximum-power current $V_i/(2r)$,
\begin{equation}
\label{eq:tau-min-power}
\tau_{\min}^{P}
=
\sum_{i=1}^{N}
\frac{C(V_{i-1}-V_i)}{V_i/(2r)} .
\end{equation}
{If} $\tau_{\max}<\tau_{\min}^{P}$, the~discharge is infeasible. As~$\tau_{\max}$ increases, $I_0$ decreases, so the solution moves from the single-current region, where all stages share the same current, to~the demand-limited region, where stages with $I_i^->I_0$ remain at their lower bounds and the remaining stages share $I_0$.

This behavior is summarized in Figure \ref{fig:feasibility-bounds}{b}. 
The map uses the five-stage load profile introduced in \cref{sec:mscd-opt},
with its relative shape fixed and only the overall power amplitude scaled. The~vertical axis is the largest stage-power ratio $\max_i P_i^{\req}/P_{\max,i}$, and~the horizontal axis is the deadline tightness $\tau_{\min}^{P}/\tau_{\max}$, where $P_{\max,i}=V_i^2/(4r)$. The~curve $I_0=\max_i I_i^-$ separates the demand-limited MSCD region from the single-current region. Above~the curve, at~least one stage has $I_i^->I_0$, and, therefore, remains at its lower bound, while the other stages share $I_0$. Below~the curve, $I_0>\max_i I_i^-$, so all
stages share one current. The~green demand-limited region is, therefore, the part of the parameter space where MSCD reduces heat relative to the peak-current~benchmark.

Together, the~two panels show the same KKT separation applied to two types of current bounds: the efficiency constraint sets upper bounds, while the load demands set lower bounds. In~the following section, we focus on the load-constrained problem with prescribed stage powers $P_i^{\req}$ and use Equation~\eqref{eq:Istar} to construct the $Q$--$\tau$ Pareto~front.

\subsection{Internal Resistance and the Three-Dimensional~Surface}\label{sec:3d}

Sweeping $\tau_{\max}$ over the feasible part of $[10,40]\,\mathrm{min}$ in Equation~\eqref{eq:I0-eq} and substituting the result into Equation~\eqref{eq:QMSCD} traces the $Q$--$\tau$ Pareto front: the set of schedules for which heat cannot be reduced without increasing the discharge time~\cite{Miettinen1999}. Shorter deadlines force a larger $I_{0}$ and produce more heat; longer deadlines reduce $I_0$, so the schedule approaches the stage-wise lower bounds $I_i^-$ and the dissipation decreases. A~knee point, therefore, provides a convenient representative compromise between time and heat. Normalizing discharging time and dissipation, we define
\begin{equation}\label{eq:normalize}
x(\tau)=
\frac{\tau-\tau_{\mathrm{scan}}^{\min}}
{\tau_{\mathrm{scan}}^{\max}-\tau_{\mathrm{scan}}^{\min}},
\qquad
y(\tau)=\frac{Q(\tau)-Q_{\min}}{Q_{\max}-Q_{\min}},
\end{equation}
where $\tau_{\mathrm{scan}}^{\min}$ and $\tau_{\mathrm{scan}}^{\max}$ are the lower and upper endpoints of the deadline scan, and~$Q_{\min}$ and $Q_{\max}$ are the corresponding endpoint values of $Q$. The~knee point is taken as the point of maximum perpendicular distance from the secant $x+y=1$ joining the two endpoints~\cite{Satopaa2011},
\begin{equation}
\label{eq:knee}
\tau_c=\operatorname*{arg\,max}_{\tau}
\frac{|x(\tau)+y(\tau)-1|}{\sqrt{2}} .
\end{equation}

We now return to the five-stage example introduced in \cref{sec:mscd-opt} and use it to quantify the knee point of the $Q$--$\tau$ Pareto front. Sweeping $\tau_{\max}$ over the feasible part of $[10,40]\,\mathrm{min}$ in Equation~\eqref{eq:I0-eq} and substituting the resulting currents into Equation~\eqref{eq:QMSCD} gives the front. Applying the knee criterion gives $\tau_c\simeq21.94\,\mathrm{min}$, $Q_c\simeq10.75\,\mathrm{kJ}$, $\eta_c\simeq77.45\%$, and~$I_0\simeq9.37\,\mathrm{A}$.  Other points on the Pareto front remain feasible and may be chosen when an application deliberately prioritizes speed or~efficiency.

\begin{figure}[H]
  \includegraphics[width=0.7\textwidth]{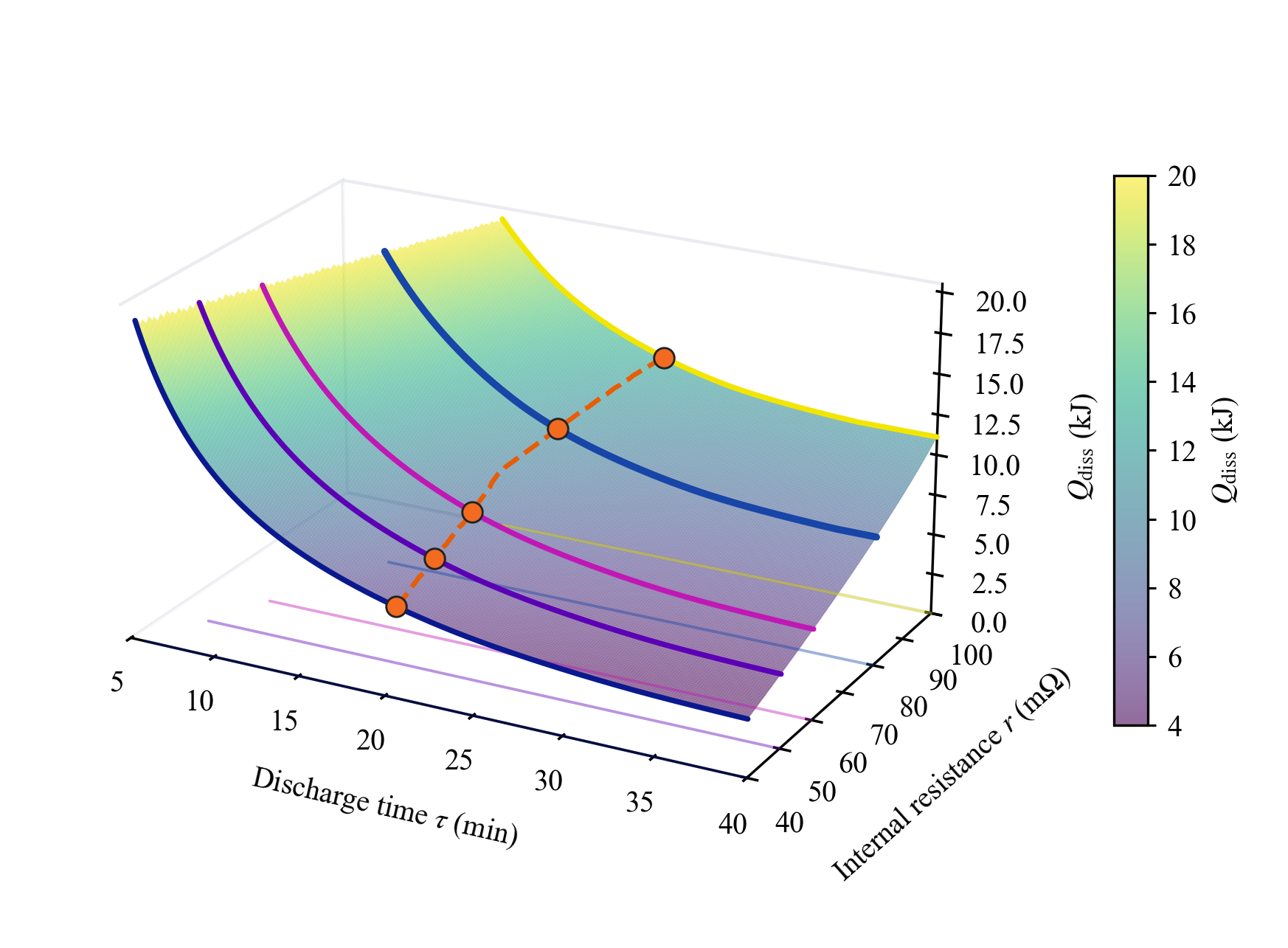}
  \caption{{Three-dimensional} 
Pareto surface $Q_{\mathrm{MSCD}}(\tau,r)$ obtained by sweeping the deadline for $r\in[40,100]$~m$\Omega$, using $P_i^{\req}=\{35.71,32.14,20.09,16.07,8.04\}\,\mathrm{W}$. Color encodes $Q_{\mathrm{diss}}$. {The dark-blue, purple, magenta, blue, and yellow solid curves show the constant-resistance Pareto fronts at $r=40$, $50$, $60$, $80$, and $100~\mathrm{m}\Omega$, respectively. The orange dashed line and markers indicate the corner points of these Pareto fronts.}}
  \label{fig:3d}
\end{figure}

The preceding $Q$--$\tau$ Pareto front was obtained at a fixed internal resistance. Since the internal resistance varies among cells and can increase during ageing~\cite{Vetter2005,Wang2020Review}, we finally examine how internal resistance $r$ reshapes the trade-off. Varying $r$ over $[0.04,0.10]\,\Omega$ and solving  Equation~\eqref{eq:I0-eq} and Equation \eqref{eq:QMSCD} at each value, the~surface $Q_{\mathrm{MSCD}}(\tau,r)$ is obtained, as~shown in Figure \ref{fig:3d}. The~surface shows two trends. At~fixed discharge time $\tau$, a larger $r$ produces greater Joule heat. Each two-dimensional slice has its own knee point on the $Q$--$\tau$ front for a fixed $r$. As~$r$ increases, this knee-point locus moves toward larger $Q$ and larger $\tau$, showing that a higher-resistance cell dissipates more heat when tight deadlines must be met. Reducing $r$ lowers the whole surface and allows shorter deadlines to be reached with less~dissipation.

\vspace{-6pt}
\unskip

\section{Effect of Series~Inductance}
\label{sec:inductance-summary}

{The preceding analysis relied on resistance-dominated equivalent circuits to isolate the fundamental Ohmic-loss mechanism. To~verify the generality of the resulting power--efficiency relations and the MSCD optimization, we introduce an effective series inductance $L$ into the circuit {(}
Figure~\ref{fig:inductive-circuits}{)}. This inductance physically represents the high-frequency parasitic behaviors of cell tabs, cable connections, and~test leads~\cite{Estaller2022}. Because~an ideal inductor exclusively stores and releases magnetic energy, it modulates the transient current dynamics and converter-voltage requirements without producing additional irreversible dissipation~\cite{Tsirlin2024}.

\vspace{-6pt}
\begin{figure}[H]
\hspace{-15pt}
 \includegraphics[width=\textwidth]{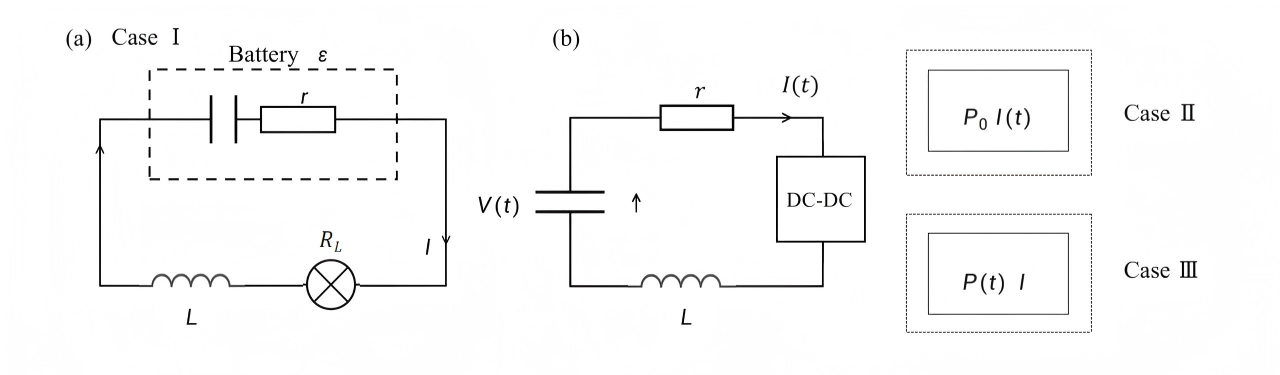}
 \caption{Equivalent circuits including an effective series inductance $L$.
(\textbf{a}) Case I: a constant-EMF battery with internal resistance $r$ and resistive load $R_L$ where $\varepsilon\equiv V_0$. (\textbf{b}) Cases II and III: a finite-capacitance battery connected to an ideal DC--DC converter. The~converter maintains constant load power $P_0$ in Case II and constant current $I$ in Case III.}
 \label{fig:inductive-circuits}
\end{figure}
}
\subsection{General Dynamic Power~Balance}

{Across all three configurations in Figure~\ref{fig:inductive-circuits}, the~instantaneous power balance incorporating the useful load power $P(t)$ is modified to:
\begin{equation}
 V(t)I(t)=P(t)+rI^2(t)+LI(t)\dot I(t).
 \label{eq:inductive-balance-main}
\end{equation}
{Here}, the~inductive term $LI\dot I=\dd(LI^2/2)/\dd t$ represents the conservative flux of magnetic energy rather than an additional irreversible loss. Consequently, the~instantaneous efficiency $\eta(t)$ and a dimensionless transient modulation factor $\alpha(t)$ are defined as:
\begin{equation}
 \eta(t)=\frac{P(t)}{V(t)I(t)}
 =1-\frac{rI(t)}{V(t)}-\frac{L\dot I(t)}{V(t)},
 \qquad
 \alpha(t)=1-\frac{L\dot I(t)}{V(t)}.
 \label{eq:inductive-efficiency-main}
\end{equation}
{Eliminating} the current $I(t)$ yields the generalized dynamic envelope:
\begin{equation}
 P(t)=\frac{V^2(t)}{r}
 \eta(t)\bigl[\alpha(t)-\eta(t)\bigr].
 \label{eq:inductive-alpha-main}
\end{equation}
{Crucially}, because~$\alpha(t)$ depends dynamically on the current trajectory via $\dot I(t)$, Equation~\eqref{eq:inductive-alpha-main} is no longer a closed algebraic relation. Specifying the transient power--efficiency trade-off and its maximum-power condition now strictly requires a predefined load model or converter-control law.
}
\subsection{Implications for the Three~Cases}
{
For Case I, the~battery drives a fixed resistive load $R_L$; thus, the current relaxes exponentially with the characteristic time $\tau_L=L/(r+R_L)$. During~the transient phase, the~output power follows $P(t)=V_0^2\eta^2(t)/R_L$, which is fixed by the load resistance. Therefore, the~inductance only determines the evolution rate of $\eta(t)$ without influencing the instantaneous algebraic relation. After~the transient decays and the inductive voltage vanishes, the~steady-state parabola $P_\infty=V_0^2\eta_\infty(1-\eta_\infty)/r$ is recovered, leaving the maximum-power efficiency $\eta_{\mathrm{EMP}}=1/2$ unchanged.}
{
For Case II, the~battery supplies a constant load power $P(t)=P_0$. The~finite capacitance imposes the kinematic constraint $C\dot V=-I$, while the inclusion of $L$ elevates the current to an independent dynamic state variable. Consequently, the~generalized envelope expands into an exact differential constraint:
\begin{equation}
 \frac{L}{CV^2}P_0^2
 +(r\eta-L\dot\eta)P_0
 -\eta^2V^2(1-\eta)=0.
 \label{eq:inductive-case-II-main}
\end{equation}
{When} the inductive transients decay ($L\to0$ or $\dot\eta\to0$), this seamlessly reduces to the purely algebraic baseline relation:
\begin{equation}
 P_0=\frac{V^2}{r}\eta(1-\eta).
 \label{eq:inductive-case-II-limit-main}
\end{equation}
{Thus}, inductance replaces the algebraic trade-off with a differential one during transients. Furthermore, a~local stability analysis reveals that under an ideal constant-power demand, the~high-efficiency (low-current) operating branch is dynamically unstable and must be actively stabilized by the converter. For Case III, the~battery operates under the multistage constant-current protocol, meaning $\dot I=0$ during each bulk discharging stage. Therefore, the~internal stage-wise power--efficiency relations remain entirely unaffected. Inductive effects are strictly confined to the finite switching intervals between stages. For~a transition $i$ of duration $\delta_i$ with a current step $\Delta I_i=I_{i+1}-I_i$, the~transient penalties scale as:
\begin{equation}
 q_i^{\mathrm{sw}}=O(\delta_i),
 \qquad
 Q_i^{\mathrm{sw}}=O(\delta_i),
 \qquad
 L\dot I=O\left(\frac{L|\Delta I_i|}{\delta_i}\right),
 \label{eq:inductive-switching-scaling}
\end{equation}
where $q_i^{\mathrm{sw}}$ and $Q_i^{\mathrm{sw}}$ are the transition charge and Joule heat. Because~a rapid current transition induces a large voltage spike $L\dot I$, the~hardware's converter-voltage limit establishes a strict lower bound on $\delta_i$. Provided $\delta_i$ remains much shorter than the bulk stage duration $T_i$, these switching penalties are negligible, robustly preserving the resistance-only MSCD constraints, dissipation functional, and~KKT solution. These results demonstrate that series inductance governs switching transients, converter-voltage limits, and~local stability, but~leaves the fundamental steady-state and constant-current thermodynamic bounds mathematically intact. Rigorous derivations, stability proofs, and~quantitative switching corrections are detailed in Appendix~\ref{app:inductance}.
}
\section{Summary and~Outlook}\label{sec:conclusion}\label{sec:outlook}

In this paper, we investigated battery discharging from a finite-time thermodynamic perspective. By~evaluating three representative models---constant electromotive force, finite-capacitance RC dynamics, and~active constant-current control---we identified a universal parabolic power--efficiency envelope, $P\propto\eta(1-\eta)$. Across all models, the~efficiency at maximum power strictly bounds to one half, serving as the exact electrochemical counterpart to the half-Carnot limit~\cite{VandenBroeck2005}. {We further demonstrated that introducing a series inductance only modifies the transient power--efficiency dynamics and converter-voltage requirements, leaving the steady-state parabola and constant-current stage relations strictly intact. Consequently, the~resistance-only MSCD formulation is seamlessly recovered provided that the switching intervals are short and remain feasible under converter-voltage limits.} Building upon these robust bounds, we formulated the multistage constant-discharging (MSCD) optimization under stage-wise load demands and a global deadline. The~Karush--Kuhn--Tucker (KKT) conditions yielded a remarkably compact optimal policy: $I_i^\star=\max(I_i^{\req},I_0)$, where the uniform baseline $I_0$ is fixed by the deadline. Conversely, a~minimum instantaneous-efficiency constraint yielded the dual upper-bound structure $I_i^\star=\min(I_c,I_i^\eta)$. These complementary solutions successfully extend the multistage charging framework~\cite{Notten2005,Zhang2014MSCC,Lee2020Optimal,Lei2025} into the load-driven discharging~regime.

For the load-constrained MSCD problem, sweeping the global deadline traces a robust dissipation--time ($Q$--$\tau$) Pareto front~\cite{Miettinen1999,Satopaa2011}. As~demonstrated in our PHEV analysis, the~MSCD strategy significantly mitigated Joule heating compared to peak-current benchmarks. We also quantified how increasing the internal resistance globally shifts this operational boundary, elevating the dissipated heat and pushing the Pareto knee point toward longer discharging times. Crucially, the~current study, together with our previous work on optimal charging~\cite{Lei2025}, establishes a complete and unified finite-time thermodynamic framework for the entire battery charge--discharge cycle. This comprehensive paradigm paves the way for analyzing multidimensional performance trade-offs in practical energy~storage.

Future work will bridge the gap between these idealized thermodynamic bounds and practical applications through several realistic extensions. {First, incorporating more detailed circuit architectures and heat-generation mechanisms---such as higher-order RC/Thevenin/Randles elements and entropy-change heat from electrode reactions~\cite{Yang2025Energy139041}---will rigorously test the robustness of the proposed power--efficiency bound beyond the minimal equivalent circuit.} Second, relaxing the assumption of a constant internal resistance by formulating it dynamically as a function of temperature and state of charge (SOC) will naturally couple the discharging dynamics to a rigorous thermal balance framework~\cite{Pesaran2002,Wang2016Thermal}. Furthermore, the~analysis can be enriched by integrating advanced electrochemical models to capture internal concentration gradients~\cite{Newman1975,Doyle1993,Sulzer2021}, alongside long-term degradation costs as competing optimization objectives~\cite{Vetter2005,Wang2020Review}. These extensions will directly empower the scheduling layer of modern Battery Management Systems while preserving the elegantly compact KKT current-bound structure established~here.

\vspace{6pt}
\section*{Author contributions}
Investigation, writing—original draft preparation, writing—review and editing, R.-H.L., Y.-Q.L. and Y.-H.M.; supervision, Y.-Q.L. and Y.-H.M.; conceptualization and funding acquisition, Y.-H.M. All authors have read and agreed to the published version of the manuscript.

\section*{Funding}
This research was funded by the National Natural Science Foundation of China under grant No.~12305037.

\section*{Data availability}
The original contributions presented in this study are included in the article. Further inquiries can be directed to the corresponding author.

\section*{Acknowledgments}
The authors thank Jia-Rui Lei for their valuable comments on this study. The~authors used {ChatGPT-5.5}
and {Gemini 3Pro} for language polishing and code-debugging assistance. All scientific content, numerical implementation, and~interpretation of results were checked and finalized by the~author.

\section*{Conflicts of interest}
The authors declare no conflicts of~interest.

\appendix
\section[\appendixname~\thesection]{Detailed Analysis of Series~Inductance}
\label{app:inductance}

This appendix details the derivations, stability proofs, numerical validations, and~MSCD switching corrections supporting Section~\ref{sec:inductance-summary}. The~overarching dynamic power balance is given by Equation~\eqref{eq:inductive-balance-main}.

\subsection[\appendixname~\thesubsection]{Case I: Transient and Steady-State~Relations}

For a constant-EMF source $V_0$ driving a resistive load $R_L$, the~circuit equation $L\dot I+(r+R_L)I=V_0$ yields the exponential relaxation
\begin{equation}
 I(t)=I_\infty+[I(0)-I_\infty]e^{-t/\tau_L},
 \label{eq:app-RL-solution}
\end{equation}
where $I(0)$ is the initial current, $I_\infty=V_0/(r+R_L)$ is the steady-state current, and~$\tau_L=L/(r+R_L)$ is the inductive relaxation time. For~a fixed $R_L$, the~instantaneous power and efficiency are $P(t)=R_LI^2(t)$ and $\eta(t)=R_LI(t)/V_0$. Eliminating $I(t)$ directly reveals
\begin{equation}
 P(t)=\frac{V_0^2}{R_L}\eta^2(t).
 \label{eq:app-RL-transient}
\end{equation}
{Thus}, $L$ merely dictates the temporal relaxation rate $\tau_L$, leaving the geometric trajectory of the transient power--efficiency relation completely unaltered. As~$t\to\infty$, the~inductive voltage $L\dot I\to0$. Eliminating $R_L$ from the asymptotic states seamlessly recovers the steady-state parabola:
\begin{equation}
 P_\infty=\frac{V_0^2}{r}
 \eta_\infty(1-\eta_\infty).
 \label{eq:app-steady-parabola}
\end{equation}

\subsection[\appendixname~\thesubsection]{Case II: Differential Constraint and Local~Stability}
\label{app:case2-stability}

Under the finite-capacitance kinematic constraint $C\dot V=-I$, substituting $I=P/(\eta V)$ into Equation~\eqref{eq:inductive-balance-main} elevates $I(t)$ to an independent dynamic state:
\begin{equation}
 \dot\eta=
 \eta\frac{\dot P}{P}
 +\frac{P}{CV^2}
 +\frac{r\eta}{L}
 -\frac{\eta^2V^2(1-\eta)}{LP},
 \qquad
 \dot V=-\frac{P}{C\eta V}.
 \label{eq:app-eta-dynamics}
\end{equation}
{For} a strictly constant load power $P(t)=P_0$, the~system simplifies to the exact differential constraint
\begin{equation}
 \frac{L}{CV^2}P_0^2+(r\eta-L\dot\eta)P_0
 -\eta^2V^2(1-\eta)=0,
 \label{eq:app-dynamic-constraint}
\end{equation}
which smoothly degenerates to the algebraic baseline $P_0=\frac{V^2}{r}\eta(1-\eta)$ as $L\to0$. Crucially, the~total energy balance satisfies
\begin{equation}
 \frac{\dd}{\dd t}
 \left(\frac{1}{2}CV^2+\frac{1}{2}LI^2\right)
 =-P(t)-rI^2,
 \label{eq:app-energy-law}
\end{equation}
confirming that inductance exclusively governs reversible magnetic-energy storage without injecting additional irreversible~dissipation.

To evaluate local stability under $P(t)=P_0$, we fix $V$ at its instantaneous value. The~current subsystem admits two stationary roots:
\begin{equation}
 I_\pm(V)=\frac{V\pm\sqrt{V^2-4rP_0}}{2r},
 \qquad
 0<P_0<\frac{V^2}{4r}.
 \label{eq:app-current-roots}
\end{equation}
{Writing} the current dynamics as $\dot I=f(I;V)=-\frac{r}{LI}(I-I_-)(I-I_+)$ yields the Jacobians
\begin{equation}
 f'(I_-)=\frac{\sqrt{V^2-4rP_0}}{LI_-}>0,
 \qquad
 f'(I_+)=-\frac{\sqrt{V^2-4rP_0}}{LI_+}<0.
\end{equation}
{Since} $f'(I_-)>0$, the~high-efficiency (low-current) branch $I_-$ is locally dynamically unstable under an ideal constant-power demand. Consequently, practical operation along this efficient branch must be actively stabilized by the~converter.

\subsection[\appendixname~\thesubsection]{Numerical Illustration of Controlled Active~Tracking}

To demonstrate this active stabilization, we introduce a proportional controller tracking the optimal reference $I_{\mathrm{ref}}(V)=I_-(V)$:
\begin{equation}
 L\dot I=-\kappa[I-I_{\mathrm{ref}}(V)],
 \qquad
 C\dot V=-I,
 \qquad \kappa>0.
 \label{eq:app-current-controller}
\end{equation}
{The} converter-input voltage, power, and~efficiency ratio become
\begin{equation}
 U_{\mathrm{in}}=V-rI-L\dot I,
 \qquad
 P_{\mathrm{in}}=U_{\mathrm{in}}I,
 \qquad
 \eta_{\mathrm{in}}=\frac{P_{\mathrm{in}}}{VI}.
 \label{eq:app-converter-input}
\end{equation}
{Assuming} an ideal lossless converter equipped with an energy buffer $E_{\mathrm{dc}}$, the~external load can rigidly maintain $P_0$ by absorbing the transient tracking mismatch $\dot E_{\mathrm{dc}}=P_{\mathrm{in}}-P_0$, yielding the global balance $\frac{\dd}{\dd t}(\frac{1}{2}CV^2+\frac{1}{2}LI^2+E_{\mathrm{dc}})=-P_0-rI^2$.

\begin{figure}[H]
 \includegraphics[width=0.66\textwidth]
 {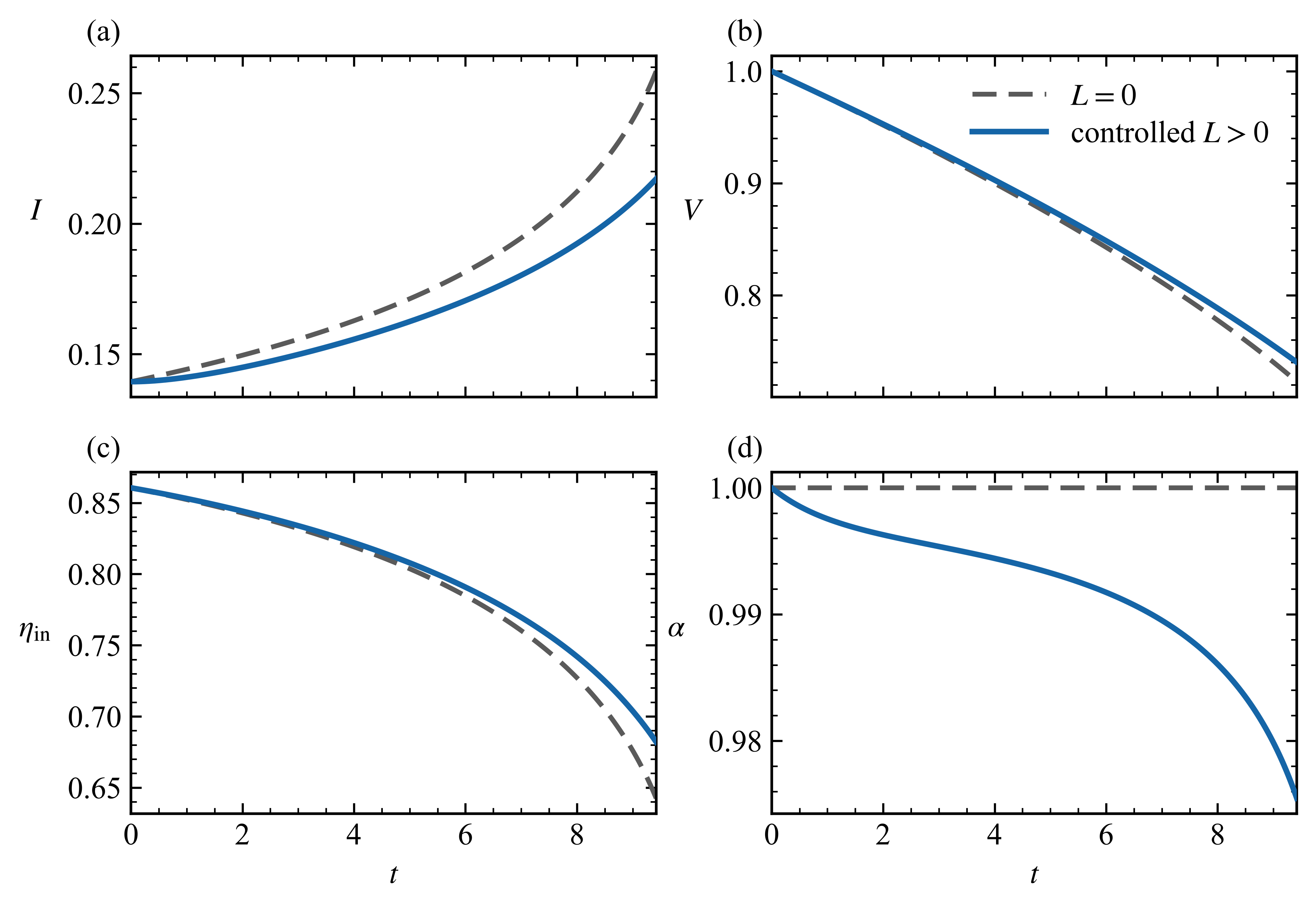}
 \caption{Controlled tracking of the low-current branch $I_-(V)$:
 (\textbf{a}) current, (\textbf{b}) source voltage, (\textbf{c}) cell-to-converter input power ratio
 $\eta_{\mathrm{in}}$, and~(\textbf{d}) transient factor $\alpha=1-L\dot I/V$.
 Dashed and solid curves denote the resistance-only limit $L=0$ and the
 controlled inductive trajectory $L>0$, respectively.}
 \label{fig:app-low-branch}
\end{figure}

Figure~\ref{fig:app-low-branch} compares the resistance-only trajectory ($L=0$) against the controlled inductive trajectory ($L>0$) using the dimensionless parameters $V(0)=1$, $r=1$, $C=6$, and $P_0=0.12$ and~deliberately large inductive/gain terms $L=0.8, \kappa=0.8$ to amplify visibility. As~$V$ decays, the~inductive lag confines the controlled current strictly below the $L=0$ limit, decelerating the source voltage drop. Because~$\dot I>0$, magnetic energy accumulates, locking the transient factor $\alpha=1-L\dot I/V$ firmly below unity. (For this specific illustrative configuration, the~maximum tracking error is $\max_t|I-I_{\mathrm{ref}}|=2.27\times10^{-2}$, with~a peak input-power mismatch of $8.64\%$.)

\vspace{-6pt}
\unskip

\subsection[\appendixname~\thesubsection]{Case III: MSCD Switching Corrections and KKT~Robustness}

During the internal constant-current stages ($\dot I=0$), the~resistance-only relations perfectly govern the MSCD protocol:
\begin{equation}
 P_i(t)=V(t)I_i-rI_i^2,
 \qquad
 \eta_i(t)=1-\frac{rI_i}{V(t)}.
 \label{eq:app-plateau-relations}
\end{equation}
{Inductive} penalties exclusively arise during the finite switching intervals between stages. To~analytically quantify this, consider a smooth cubic transition of duration $\delta_i$ linking $I_i$ and $I_{i+1}$:
\begin{equation}
 I(t)=I_i+\Delta I_i (3x^2-2x^3),
 \qquad
 x=\frac{t-t_i}{\delta_i},
 \label{eq:app-smooth-switch}
\end{equation}
which enforces $\dot I(t)=\frac{6\Delta I_i}{\delta_i}x(1-x)$ and smoothly matching boundary conditions. Integrating Eq.~\eqref{eq:app-smooth-switch} over time yields the battery voltage $V(t)$, and Kirchhoff's voltage law then determines the converter-input voltage $U_{\mathrm{in}}(t)$. The switching power $P_{\mathrm{sw}}(t)$ and instantaneous efficiency $\eta_{\mathrm{sw}}(t)$ are subsequently evaluated from the same definitions used above, with the efficiency including both ohmic and inductive voltage drops.

Assuming ideal lossless conversion, transition $i$ remains physically feasible strictly if $U_{\min}\leq U_{\mathrm{in}}(t)\leq U_{\max}$ and $P_{\mathrm{sw}}(t)\geq P_{\mathrm{req}}(t)$. Since the peak inductive spike scales as $\max_t|\dot I(t)|=3|\Delta I_i|/(2\delta_i)$, the~absolute hardware converter-voltage limits dictate the minimum allowable switching time $\delta_i$. 

Integrating the cubic profile extracts the transferred charge $q_i^{\mathrm{sw}}$ and root-mean-square switching current $I_{i,\mathrm{rms}}^{\mathrm{sw}}$:
\begin{equation}
 q_i^{\mathrm{sw}}
 =\frac{\delta_i}{2}(I_i+I_{i+1}),
 \qquad
 \left(I_{i,\mathrm{rms}}^{\mathrm{sw}}\right)^2
 =\frac{13I_i^2+9I_iI_{i+1}+13I_{i+1}^2}{35}.
 \label{eq:app-switch-charge-rms}
\end{equation}
{The} switching Joule heat and magnetic-energy shift directly follow as $Q_i^{\mathrm{sw}}=r(I_{i,\mathrm{rms}}^{\mathrm{sw}})^2\delta_i$ and $\Delta E_{L,i}=\frac{L}{2}(I_{i+1}^2-I_i^2)$. Regardless of the precise profile shape, any smooth transition yields the rigorous scaling:
\begin{equation}
 q_i^{\mathrm{sw}}=O(\delta_i),
 \qquad
 Q_i^{\mathrm{sw}}=O(\delta_i),
 \qquad
 L\dot I=O\left(\frac{L|\Delta I_i|}{\delta_i}\right).
 \label{eq:app-switch-scaling}
\end{equation}
{Accounting} for these $N-1$ finite transitions explicitly modifies the global MSCD Joule heat objective, deadline, and~charge constraints:
\begin{align}
 Q_{\mathrm{tot}}&=r\sum_{i=1}^{N}I_i^2T_i+\sum_{i=1}^{N-1}Q_i^{\mathrm{sw}}, \label{eq:app-corrected-objective} \\
 \tau_{\max} &\geq \sum_{i=1}^{N}T_i+\sum_{i=1}^{N-1}\delta_i, \label{eq:app-corrected-time} \\
 C(V_0-V_f) &= \sum_{i=1}^{N}I_iT_i+\sum_{i=1}^{N-1}q_i^{\mathrm{sw}}. \label{eq:app-corrected-charge}
\end{align}
{Because} $q_i^{\mathrm{sw}}$ and $Q_i^{\mathrm{sw}}$ dynamically couple adjacent stages $I_i$ and $I_{i+1}$, the~idealized stage-wise KKT decoupling $I_i^\star=\max(I_i^{\mathrm{req}},I_0)$ is no longer strictly exact. However, provided the switching intervals remain brief relative to the bulk stages ($\delta_i\ll T_i$) and feasible under converter limits, the~switching penalties are completely overshadowed: $\sum q_i^{\mathrm{sw}} \ll \sum I_iT_i$ and $\sum Q_i^{\mathrm{sw}} \ll r\sum I_i^2T_i$. Under~this practical operational limit, the~finite-switching corrections vanish, robustly restoring the exact resistance-only MSCD optimization~structure.

\bibliographystyle{apsrev4-2}
\bibliography{references}
\end{document}